\documentclass[aps,10pt,pre,twocolumn,nofootinbib,tightenlines,superscriptaddress,amsmath,amssymb,final,letterpaper,longbibliography]{revtex4-1}
\usepackage[utf8]{inputenc}
\usepackage{graphicx}
\usepackage{color}
\definecolor{seaborn_deep_blue}{RGB}{76, 114, 176}

\usepackage{hyperref}
\hypersetup{unicode=false, pdftoolbar=true, pdfmenubar=true, pdffitwindow=false, pdfstartview={FitH}, pdfnewwindow=true, colorlinks=true, allcolors=seaborn_deep_blue,}

\begin{document}
\title{Exact and rapid linear clustering of networks with dynamic programming}

\author{Alice Patania}
\email{alice.patania@uvm.edu}
\affiliation{Department of Mathematics and Statistics, University of Vermont, Burlington, VT 05405, USA}
\affiliation{Vermont Complex Systems Center, University of Vermont, Burlington, VT 05405, USA}

\author{Antoine Allard}
\affiliation{Vermont Complex Systems Center, University of Vermont, Burlington, VT 05405, USA}
\affiliation{D\'epartement de physique, de g\'enie physique et d'optique, Universit\'e Laval, Qu\'ebec (Qu\'ebec), Canada G1V 0A6}
\affiliation{Centre interdisciplinaire en mod\'elisation math\'ematique, Universit\'e Laval, Qu\'ebec (Qu\'ebec), Canada G1V 0A6}

\author{Jean-Gabriel Young}
\email{jean-gabriel.young@uvm.edu}
\affiliation{Department of Mathematics and Statistics, University of Vermont, Burlington, VT 05405, USA}
\affiliation{Vermont Complex Systems Center, University of Vermont, Burlington, VT 05405, USA}
\affiliation{D\'epartement de physique, de g\'enie physique et d'optique, Universit\'e Laval, Qu\'ebec (Qu\'ebec), Canada G1V 0A6}

\begin{abstract}
We study the problem of clustering networks whose nodes have imputed or physical positions in a single dimension, for example prestige hierarchies or the similarity dimension of hyperbolic embeddings.
Existing algorithms, such as the critical gap method and other greedy strategies, only offer approximate solutions to this problem.
Here, we introduce a dynamic programming approach that returns provably optimal solutions in polynomial time---$O(n^2)$ steps---for a broad class of clustering objectives.
We demonstrate the algorithm through applications to synthetic and empirical networks, and show that it outperforms existing heuristics by a significant margin, with a similar execution time.
\end{abstract}

\maketitle
\section{Introduction}
Many complex networks live in ultra low-dimensional spaces~\cite{almagro2022detecting}.
Spatial networks are an obvious example since their nodes are situated in two or three dimensions---locations on a map  or the voxels of a connectome, say---and their edges denote relationships such as geographical proximity or physical connections---cables, roads, axons, and so on~\cite{barthelemy2011spatial}.
But it also makes sense to think of several other types of networks as low-dimensional.
For example, the nodes of social networks can often be ordered according to social constructs such as ``power''~\cite{pinter2014dynamics}, ``reputation''~\cite{power2018building}, ``attractiveness''~\cite{bruch2018aspirational} or
even ``centrality''~\cite{gleich2015pagerank},  and as a result, we can think of their nodes as embedded along a single dimension---or perhaps a few more if we use more than one property simultaneously.
And, of course, networks that capture more abstract connections, such as social relationships in a school~\cite{moody2001race} or the hyperlinks of the World Wide Web~\cite{page1999pagerank} can also be placed in such low-dimensional spaces by using one of several dozens of popular embedding algorithms from spectral methods~\cite{belkin2001laplacian} to graph learning~\cite{xu2021understanding}, graph layouts~\cite{diaz2002survey}, or network geometry~\cite{boguna2021network,chami2019hyperbolic}.

It is well understood that low-dimensional embeddings can help us find communities---groups of similar nodes---efficiently.
Good embeddings place similar nodes together and thus recast the problem of finding communities to a problem of splitting the network into sets of close-by nodes.
In fact, this insight has led to several popular and classical community detection that proceeds in two steps, first an embedding step that assigns coordinates to nodes, and second a clustering step that finds clusters of nodes with similar coordinates~\cite{ng2001spectral}.
A few examples of classical algorithms that work this way include spectral modularity maximization~\cite{white2005spectral,newman2006modularity}, where one embeds the nodes with the leading eigenvectors of the modularity matrix, followed by a $k$-means or recursive clustering of these embedding, or any of several variations on graph clustering with the Laplacian matrix~\cite{riolo2014first,ng2001spectral}.
Most modern graph machine learning pipelines follow a similar two steps process~\cite{xu2021understanding}.

The main contribution of this paper is a fast method for finding optimal contiguous communities when the nodes have positions in a \emph{single} dimension---the linear clustering problem.
The algorithm takes an embedded network and an objective function as input, and returns an optimal partition in as few as $O(n^2)$ steps where $n$ is the number of nodes, independent from the number of communities $q$ in the partition.
This is faster than a brute-force search as soon as $q>2$ or when $q$ is unknown.
We achieve this result by  adapting well-known dynamic programming ideas to linear clustering for networks~\cite{bellman2013dynamic,jackson2005algorithm}.

Our algorithm nonetheless solves two problems of current interest.
One, recent work has focused on developing ranking algorithms for networks~\cite{ball2013friendship,de2018physical,cantwell2022belief,ragain2016pairwise,newman2022rankings}, and raised the possibility that nodes may naturally split into layers, i.e., largely disconnected groups of nodes with different average ranks~\cite{de2018physical,kawamoto2023consistency}.
Some rankings are even designed with this goal in mind~\cite{jokic2023linear,shang2022local,ochi2022finding}.
Two, a different line of work has revealed that many of the properties of the observed empirical network can be explained by thinking of the nodes as embedded in a very low-dimensional hyperbolic space~\cite{boguna2021network,almagro2022detecting,osat2023embeddingaided}.
Crucially, when the embedding space is a circle, the angular coordinate of nodes tends to correlate with communities found by algorithms such as modularity maximization~\cite{tandon2021community}.
In both problems, recent work has focused largely on the embedding algorithms themselves and has left the subsequent partitioning step to greedy maximization techniques such as the iterative aggregation or the critical gap method~\cite{tandon2021community,serrano2012uncovering,bruno2019community}.
As we will show, these greedy algorithms can sometimes perform poorly; a fast algorithm for partitioning the dimension would greatly simplify the analysis.

\begin{figure*}
  \centering
  \includegraphics[width=0.63\linewidth]{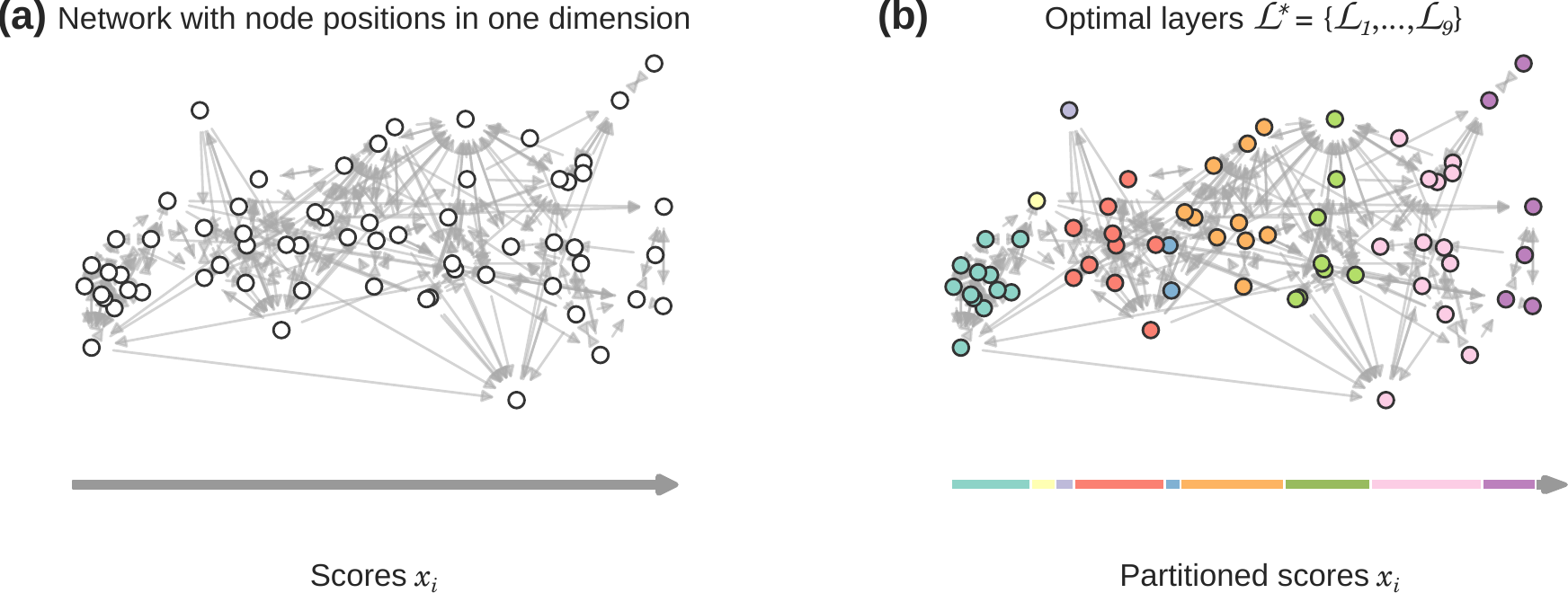}
  \caption{\textbf{The linear clustering problem}. \textbf{(a)}~The problem's input is a network whose nodes have positions in one dimension. These positions can be observed covariates---for example, wealth, ranking, or caste---or they can be estimated with latent space models, graph embedding methods, or other methods. \textbf{(b)} Our algorithm returns a provably optimal partition of the nodes in contiguous communities, called layers $\mathcal{L}=\{L_1,...L_q\}$, for any objective functions of the form $Q(\mathcal{L}) = \sum_{r=1}^q f(L_r)$ where $f(L_r)$ is an arbitrary function of the nodes in layer $L_r$.}
  \label{fig:cartoon}
\end{figure*}

The paper is organized as follows.
We state the problem formally in Sec.~\ref{sec:problem} and explain the dynamic program in Sec.~\ref{sec:method}.
We then demonstrate its use in three case studies regrouped in Sec.~\ref{sec:results}, namely a simulation study demonstrating quality and computational improvements brought about by the algorithm; the detection of social strata in status hierarchies; and the detection of communities using positions revealed by a hyperbolic embedding. 
Conclusions follow in Sec.~\ref{sec:conclusion}.
A reference \texttt{python} implementation of the algorithm is available online at \url{github.com/jg-you/dyvider}.

% ~~~~~~~~~~~~~~~~~~~~~~~~~~~~~~~~~~~~~~~~~~~~~~~
\section{The problem}
\label{sec:problem}
% ~~~~~~~~~~~~~~~~~~~~~~~~~~~~~~~~~~~~~~~~~~~~~~~

Consider a network whose nodes have a fixed position in a one-dimensional space.
For the sake of concreteness, we will think of these positions as ``scores'' though any notion of position could replace these scores, say a discrete or continuous network centrality, covariates such as wealth, attractiveness, and status, or an embedding found with a spectral method or a graph neural network~\cite{diaz2002survey}.
Without loss of generality, let us encode the score of a node $i$ as a real number $x_i$, satisfying $x_i > x_j$ if node $i$ has a higher score than node $j$.

\subsection{Linear clustering}
In this paper we will solve the task of partitioning these nodes in \emph{contiguous} communities, i.e., communities such that if three nodes $i,j,$ and $k$ have decreasing scores $x_i > x_j > x_k$, then nodes $i$ and $k$ can be in the same community if and only if $j$ is also part of that community.

Since contiguity differentiates these groups of nodes from the more classical notion of community, we will call these communities layers and denote them as $\mathcal{L}=\{L_1,...,L_q\}$ where $L_r$ is the $r$\textsuperscript{th} set of contiguous nodes; see Fig.~\ref{fig:cartoon}.

This \emph{linear clustering} task can be formalized as a discrete optimization problem, in which an objective function $Q(\mathcal{L})$ tells us how ``good'' any given partition is.
The objective may reflect various desirable properties of the layers; for example, it could increase with the internal density of layers.
The problem can then be solved as a search over all possible partitions of the nodes in layers $\mathcal{L}$ guided by $Q$.

It is worth highlighting that even though this paper introduces a rapid and exact algorithm for the linear clustering problem, the problem is not obviously easy from a computational perspective.
Indeed, despite the structure imposed by the contiguity constraint, a brute-force search enumerating all solutions is prohibitively costly.
For a fixed number of layers $q$ there are
\begin{equation}
  \Omega(n,q)=\binom{n-1}{q-1} = O(n^{q-1})   
\end{equation} 
possible contiguous partitions $\mathcal{L}$, namely the number of ways of splitting the scores with $q-1$ cut points falling in the $n-1$ empty spaces between the $n$ nodes.
While $\Omega(n,q)$ is small when $q=2$, empirical systems can have dozens of layers  and thousands if not millions of nodes.
Further, for an unknown number of layers $q$, the size of the solution space grows exponentially since $\sum_{q=1}^{n}\binom{n-1}{q-1} = 2^{n-1}$.
Hence, a search is not realistically feasible in all but the simplest cases.

In Sec.~\ref{sec:method} below, we introduce  a dynamic programming (DP) algorithm that can find the optimal layers  efficiently.
We observe the classic efficiency gains of DP and are thus able to maximize $Q$ over the exponentially large solution space in as few as $O(n^2)$ steps. 

For the proposed approach to work, however, we will need to constraint the types of allowed objective $Q$ and thus consider objective functions that give the quality of a partition in layers $\mathcal{L}$ as the sum
\begin{equation}
  \label{eq:sum_constraint}
  Q(\mathcal{L}) = \sum_{r=1}^{q} f(L_r),
\end{equation}
where $f(L_r)$ is the quality of layer $L_r$. 
Not all functions can be written this way,\footnote{For example, the log-likelihood of the degree-corrected stochastic block model~\cite{karrer2011stochastic}.} but it turns out that several standards, interesting, or otherwise natural objective functions have this form.

% ~~~~~~~~~~~~~~~~~~~~~~~~~~~~~~~~~~~~~~~~~~~~~~~
\subsection{Objective functions for layers}
\label{sec:objective}
% ~~~~~~~~~~~~~~~~~~~~~~~~~~~~~~~~~~~~~~~~~~~~~~~

Perhaps the most common application of graph clustering is to find groups of highly connected nodes.
The quality of a set $\mathcal{L}$ of layers under this definition can be captured as
\begin{equation}
  \label{eq:unpenalized}
  Q(\mathcal{L}) = \sum_{r=1}^{q}m_r 
\end{equation} 
where $m_r$ is the number of edges connecting pairs of nodes.
This objective manifestly has the same form as Eq.~\eqref{eq:sum_constraint} and is thus a first example of objectives that can be handled in our framework.

But Eq.~\eqref{eq:unpenalized} is not very interesting, in the sense that one can trivially maximize the number of internal edges by creating a large layer that contains (nearly) all the nodes and edges---thereby saying nothing of substance about the graph itself.
Hence, one usually introduces a penalty / regularization term to  rule out uninteresting solutions.
As long as this penalty can be written as a sum over layers, say a quadratic penalty $\sum_{r=1}^qn_r^2$ where $n_r$ is the number of nodes in layer $r$, then the resulting objective function will correspond to the summation form in Eq.~\eqref{eq:sum_constraint}, e.g.,
\begin{equation}
  Q(\mathcal{L}) = \sum_{r=1}^q   (m_r - n_r^2),
\end{equation} 
with layer qualities $f(L_r)=m_r-n_r^2$.

Modularity-like metrics, which compare the number of edges inside layers to some null hypothesis, are another natural example.
Using  $A_{ij}\in\{0,1\}$ to denote an entry of the adjacency matrix and $l_i\in\{1,...,q\}$ to encode the layer of node $i$, we define the modularity as
\begin{equation}
  \label{eq:modularity}
  Q(\mathcal{L}) = \frac{1}{2m}\sum_{i=1}^n\sum_{j=1}^n   \left[A_{ij} - P_{ij} \right] \delta_{l_il_j},
\end{equation}
where $m$ is the number of edges,  $\delta_{rs}$ is the Kronecker delta, and $P_{ij}$ is the probability of an edge under some null model.
In the context of networks embedded in one dimension, this null model could depend on the scores $\vec{x}$ or be the more traditional configuration model
\begin{equation}
  \label{eq:cm_null}
  P_{ij} = \frac{k_ik_j}{2m}
\end{equation}
where $k_i$ is the degree of node $i$.
Regardless of the choice of $P_{ij}$, one can re-arrange Eq.~\eqref{eq:modularity} and calculate the modularity as the sum over layers
\begin{equation}
  Q(\mathcal{L}) = \frac{1}{m}\sum_{r=1}^q \Big(m_r - \langle m_r\rangle\Big) ,
\end{equation}
thus recovering Eq.~\eqref{eq:sum_constraint}.
In this modularity, we interpret $\langle m_r\rangle$ as the expectation of the number of edges in layer $r$, $m_r$, under the null model $P$.
It is calculated as $\langle m_r\rangle= \frac{1}{2}\sum_{i,j} P_{ij}\delta_{l_ir}\delta_{l_jr}$.

Finally,  one may use Eq.~\eqref{eq:sum_constraint} as a template to devise new objective functions that reflect specific theoretical constructs.
For example, in a directed network where unreciprocated edges encode a differential in social status and reciprocated edges encode equality in status, we may theorize that individuals with a similar status cluster in social strata and that layers thus contain a lot of reciprocated ties---an ``egalitarian'' formulation of social strata.
Using $A_{ij}$ to denote the presence or absence of edges from $i$ to $j$, we could then capture this construct with the objective:
\begin{equation}
  \label{eq:obj_egalitarian}
  f_E(L_r) = \sum_{(i\leq j) \in L_r} \bigl[A_{ij}A_{ji} - \varepsilon\bigr]
\end{equation}
where the sum runs over pairs of nodes in $L_r$ with replacement, and where $\varepsilon$ is introduced to penalize large communities.
(We set $\varepsilon$ using the constraint $Q=0$ when all nodes are in a single layer, leading to $\varepsilon=2w/(n(n+1))$ where $w=\sum_{i<j}A_{ij}A_{ji}$ is the number of pairs of nodes sharing reciprocated ties.)
But Eq.~\eqref{eq:obj_egalitarian} is not the only way to formalize status hierarchies.
We could instead look for social layers with as few \emph{un}reciprocated ties as possible, thereby emphasizing the lack of dominance within layers, say as
\begin{equation}
  \label{eq:obj_dominance}
  f_D(L_r) = -\sum_{(i\leq j) \in L_r} \bigl[A_{ij}(1-A_{ji}) + (1 - A_{ij})A_{ji}  - \varepsilon'\bigr],
\end{equation}
where $\varepsilon'$ is, again, a constant set using an equation for $Q=0$ to penalize trivial solutions. 
Or we could even balance the two objectives and simultaneously penalize unreciprocated edges while favoring reciprocated ones, with
\begin{equation}
  \label{eq:obj_balance}
  f_B(L_r) = \lambda f_E(L_r) + (1 - \lambda) f_D(L_r),
\end{equation}
where $\lambda\in[0,1]$ controls the emphasis on each part of the objective.
All of these different theories lead to objective functions of the form appearing in Eq.~\eqref{eq:sum_constraint} and can thus be used with the algorithm we outline next.
Statistical objectives, such as likelihood and posterior probabilities for partitions, can also be derived in the same way~\cite{Young2018}.

% ~~~~~~~~~~~~~~~~~~~~~~~~~~~~~~~~~~~~~~~~~~~~~~~
\section{Dynamic programming for linear clustering}
\label{sec:method}
% ~~~~~~~~~~~~~~~~~~~~~~~~~~~~~~~~~~~~~~~~~~~~~~~

We now describe a dynamic programming (DP) method to maximize the objective function $Q$ in Eq.~\eqref{eq:sum_constraint} in as few as $O(n^2)$ operations.
It is adapted from a generic method developed for partitioning data on an interval~\cite{jackson2005algorithm,bellman2013dynamic}, that has found many applications in computer science, from binning data optimally to optimizing the layout of \LaTeX{} documents~\cite{knuth1981breaking,moore2011nature}.
The algorithm is guaranteed to return an optimal solution, and thus, we call it an exact algorithm.

To facilitate the discussion, we label nodes by scores, such that node $1$ has score $x_1$ and subsequent nodes have non-increasing scores $x_1\geq x_2 \geq \ldots \geq x_n$.
In practice, equal scores are fairly typical---for example, scoring nodes by network centrality will yield many ties---and this can add unnecessary complications to the algorithm.
We avoid these complications by noticing that two nodes with an identical score must be placed in the same layer.
Hence, we add a pre-processing step in which we collapse all nodes with equal scores into ``super-nodes.''
After this step is completed, we may assume without loss of generality that the scores are decreasing $x_1 > ... > x_n$ for some number of \emph{super-}nodes $n$, henceforth referred to simply as the number of nodes $n$.
The network may contain self-loops and multiedges induced by the pre-processing steps or the original data.
With this ordering in place, we then use $L_{i,j}$ to refer to a layer that contains node $i$ through $j \geq  i$ inclusively, and $f(L_{i,j})$ to refer to its quality.

The algorithm is more easily understood by first focusing on the numerical value $Q^*$ of the quality of the best partition while---perhaps unintuitively---setting the actual separation in layers to the side.
The DP approach sets up this calculation by introducing $Q_j^*$, the maximum of $Q$ when optimized over possible separations $\mathcal{L}$ of the $j$ nodes with the highest scores while ignoring the bottom $n-j$ nodes.
In this notation,  $Q_n^*$ is the maximum of $Q$ for the whole network, and an algorithm that can compute $Q^*_j$ for any $j$ will thus give us the numerical value associated with the optimal separation in layers.

We compute $Q_j^*$ by recursing on $j$.
To set up the recursion, we assume that we already know the maxima $Q^*_1, Q^*_2,..., Q_{k-1}^*$ of the objective function when optimized over partitions of the first $k-1$ nodes (the $k-1$ nodes with the highest scores).
The main necessary insight is then that $Q_j^*$ can be calculated as
\begin{equation}
    \label{eq:best_split_quality}
    Q_j^* = \max_{k=1,...,j} \Big\{ Q_{k-1}^* + f(L_{k, j})\Big\},
\end{equation}
since the quality of a set of layers is the sum of their qualities, see Eq.~\eqref{eq:sum_constraint}.
The bracketed term in Eq.~\eqref{eq:best_split_quality} is the quality of a partition obtained as the union of a bottom layer comprising nodes $k$ through $j$, and some optimal set of layers of quality $Q_{k-1}^*$ for the top $k-1$ nodes.
The maximization runs over all such extensions and thus returns the quality of the optimal partition for the first $j$ nodes.
To complete the specification of the recursion, we establish that the base case is $Q_0^* = 0$.

In practice, we can unfold the recursion by calculating the solution from $j=1$ to $n$, first using $Q_{1}^*$ to compute $Q_2^*$, then using $\{Q_1^*, Q_2^*\}$ to compute $Q_3^*$, and so on.
This complete algorithm can be executed in $O(n^2)$ steps because the costliest step---the maximization in Eq.~\eqref{eq:best_split_quality}---takes $O(j)$ operations, is repeated for $j=1,...,n$, and each repetition involves a call to $f(\cdot)$.
The call to $f(\cdot)$ can, in principle, lead to worse-than-quadratic scaling for the complete algorithm if the number of steps needed to evaluate $f(\cdot)$ is a function of $n$. 
As it turns out, this function can be evaluated with an algorithm whose amortized complexity is $O(1)$ when implemented with some bookkeeping and restrictions on $f(L_{k,j})$; see Appendix~\ref{appendix:improved_implementation} for details.
Hence, the overall algorithm can be executed in a quadratic number of steps for most choices of the objective function, including all the ones considered in Sec.~\ref{sec:objective} above.

As we have alluded to previously, the above algorithm computes $Q_n^*$  but not the associated optimal partition 
\begin{equation}
  \label{eq:argmax}
  \mathcal{L}^* = \mathrm{argmax}_{\mathcal{L}}\Big\{Q (\mathcal{L})\Big\},
\end{equation}
which is the object we actually want.

We retrieve this partition with a small modification to the algorithm by storing the maximizing index $k$ at every evaluation of $Q_j^*$.
This provides sufficient information to determine the partition $\mathcal{L}^*$ associated with the optimal solution $Q_n^*$ by backtracking through stored indices once the calculation is over \cite{jackson2005algorithm}.
Starting from the end, the stored index $k$ immediately gives us the content of the bottom layer, which comprises node $k$ through $n$.
We then retrieve the index $k'$ that maximized $Q_k^*$ and construct the second-to-last layer, $L_{k',k}$.
We continue this process and build the optimal partition $\mathcal{L}^*$ by recursing until we reach node~$1$.

\begin{figure*}[!t]
  \centering
  \includegraphics[width=0.9\linewidth]{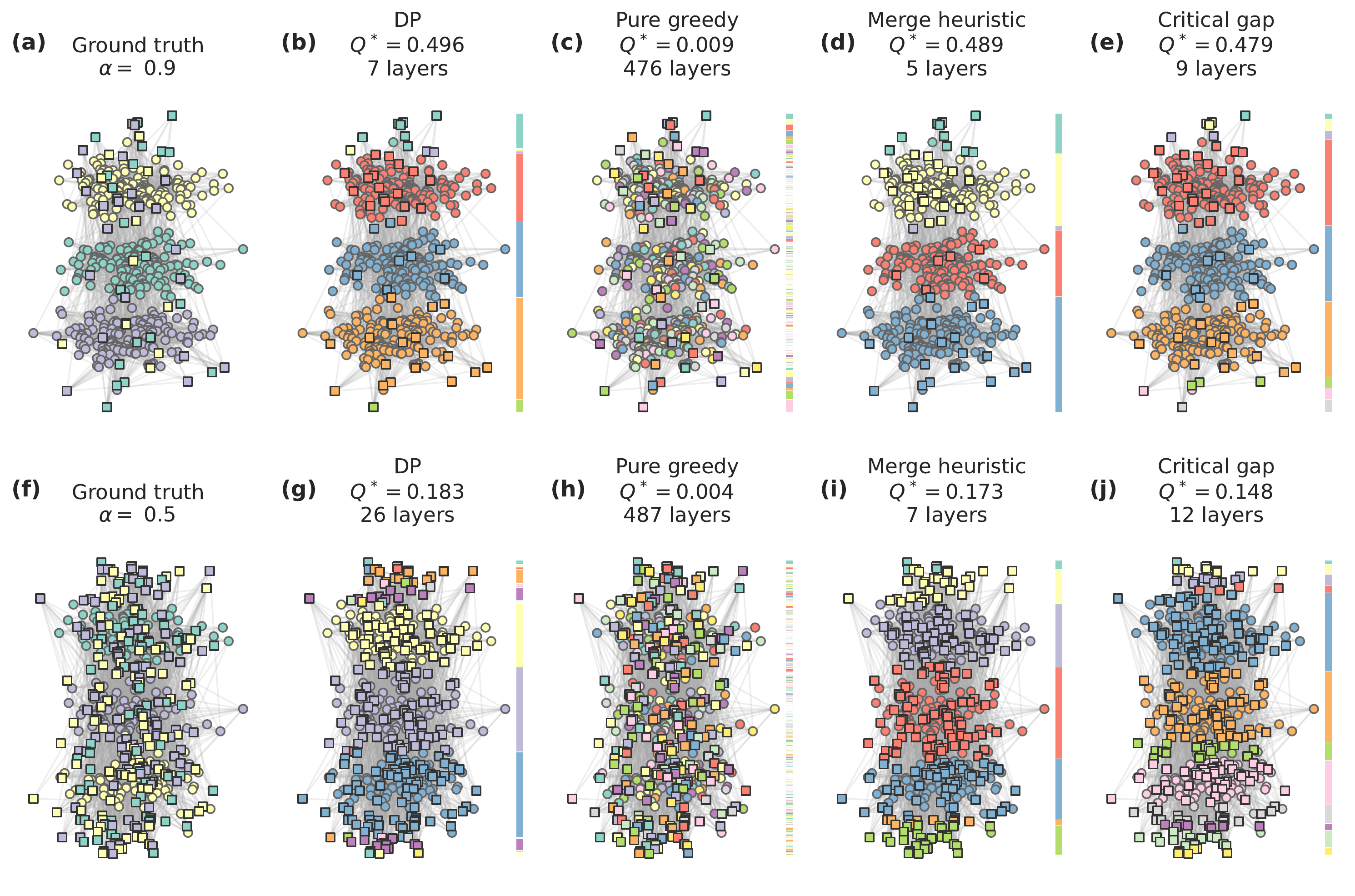}
  \caption{\textbf{Layer structure of computer-generated networks}.
  We apply linear clustering algorithms to synthetic networks whose communities (node colors and color bars) are correlated with node scores (vertical position).
  \textbf{Top row, (a-e)}. Output of the algorithms when there is a strong correlation between community labels and scores (high $\alpha$).
  The algorithms all maximize the same objective function $Q$, the standard modularity obtained by setting $P_{ij}=k_ik_j/2m$ in Eq.~\eqref{eq:modularity}, where $m$ is the number of edges in the network and $k_i$ is the degree of node $i$.
  \textbf{Bottom row, (f-j)}. Same experiment as in the top row, but for a network with little correlation between communities and scores (low $\alpha$).
  In both sets of experiments, networks have $n=500$ nodes and $k=3$ communities of roughly equal size, the standard deviation of scores within communities is $\sigma=0.05$, and the connection probabilities are $p_{\mathrm{in}}=0.05$ and $p_{\mathrm{out}}=0.0005$. 
  Each algorithm takes the scores as input and can only return contiguous layers formed by nodes with adjacent scores.
  Nodes whose score is determined by their communities are drawn with a circle, while nodes with randomized scores are drawn with squares.
  Note that all \emph{detected} communities are contiguous (b-e, g-j);  but that some distinct layers are labeled with the same color, due to a limited color palette.
  }
  \label{fig:example}
\end{figure*}

A few properties of this DP algorithm are worth highlighting.

First, it is ``exact'' in the sense that it returns a maximizing partition $\mathcal{L}^*$ of $Q$.
More precisely, there exist no contiguous partition $\tilde{\mathcal{L}}$ of strictly higher quality such that $Q(\tilde{\mathcal{L}})>Q(\mathcal{L}^*)$.

Second, the number of layers $q$ in the optimal solution $\mathcal{L}^*$ is not an input of this algorithm; $q$ is instead selected automatically.
If needed, it can be set indirectly by designing objective functions with strong constraints on the number of layers---via penalty terms, for example.

Third, we note that multiple choices of $k$ can maximize Eq.~\eqref{eq:best_split_quality} at any given point $j$ of the execution.
As a result, there are ambiguities in how the retrieval of $\mathcal{L}^*$ runs in practice, as it is unclear which index should be returned when retrieving the optimal partition if there are more than one maximizing indices.
We resolve this issue by giving priority to the smallest $k$.
Since the quality $Q_j^*$ of these optimal partial solutions is, by definition, identical for all maximizing indices $k$, the quality $Q_n^*$ of the final solution does not depend on this choice; it is only the content of the corresponding layers that may change with a different prioritization rule.
We can thus view the returned partition as one of possibly many maximizing partitions.
(The algorithm can also return all optimal partitions with a slight modification: store all maximizing indices $k$ at every $j$. 
Retrieval of the partitions then runs as a branching process, in which several maximizing indices are retrieved every time we backtrack, branching out all the way back to index 1.
This comes with some computational and storage overhead, so we focus on single solutions here.)

Fourth, while the algorithm calls for a linear embedding of the nodes, it can be applied to graphs embedded in a linear periodic space (e.g., a circle) with a few modifications.
The resulting algorithm is slightly slower but still runs in polynomial time, possibly in quadratic time even---though we only offer empirical evidence for this rather than proofs. 
Appendix~\ref{appendix:periodic} outlines the necessary modifications.

% ~~~~~~~~~~~~~~~~~~~~~~~~~~~~~~~~~~~~~~~~~~~~~~~
% ~~~~~~~~~~~~~~~~~~~~~~~~~~~~~~~~~~~~~~~~~~~~~~~
\section{Case studies}
\label{sec:results}
% ~~~~~~~~~~~~~~~~~~~~~~~~~~~~~~~~~~~~~~~~~~~~~~~
% ~~~~~~~~~~~~~~~~~~~~~~~~~~~~~~~~~~~~~~~~~~~~~~~

% ~~~~~~~~~~~~~~~~~~~~~~~~~~~~~~~~~~~~~~~~~~~~~~~
\subsection{Recovering planted layers}
\label{subsec:results_simulation_study}
% ~~~~~~~~~~~~~~~~~~~~~~~~~~~~~~~~~~~~~~~~~~~~~~~

As a first test for the algorithm, we study synthetic networks whose layer structure is explicitly correlated with the nodes' scores $\{x_i\}$.
To this end, we first generate a network with the planted partition model, by assigning each node to a layer and connecting pairs of nodes with probability $p_{\mathrm{in}}$ if they are in the same layer and with probability $p_{\mathrm{out}}<p_{\mathrm{in}}$ otherwise.
To inject information about the layers in the scores $\{x_i\}$, we then assign a central score $\mu_\ell = \ell / (q + 1)$ to each layer $\ell = 1, 2, \ldots, q$, thereby distributing them evenly on the $[0, 1]$ interval.
We finally determine the scores of each node by drawing it from a normal distribution whose mean is the central score $\mu_\ell$ of the node's layer, with a standard deviation of $\sigma$.
Finally, we control the correlation between the layer structure and scores by randomizing the score of some nodes; we replace the layer-determined score of each node  with a random score drawn uniformly from the interval $[0,1]$  with probability $1-\alpha$.
When $\alpha=0$, the scores contain no information about the layers, and they are strongly segregated by layers when $\alpha = 1$---as long as the standard deviation  $\sigma$ is not too large.
Two such networks are shown in Fig.~\ref{fig:example}~(a) and (f), the first with nearly perfect alignment of layers and scores ($\alpha=0.9$)  and the second with low alignment ($\alpha=0.5$).

The panels of Fig.~\ref{fig:example} show the partitions found by several algorithms applied to these two example networks with the standard modularity objective in Eqs.~\eqref{eq:modularity} and~\eqref{eq:cm_null}.
Alongside the partitions found with the dynamic programming (DP) method described in Sec.~\ref{sec:method}, which we display in panels (b) and (g), we show the output of a few heuristics.
The simplest heuristic is a purely greedy algorithm that starts from $n$ layers of one node and merges pairs of contiguous layers until no improvements to $Q$ can be found.
We also implement a more sophisticated merging heuristic that accepts a temporary decrease to $Q$ (if no good moves are available), runs until all layers have been merged into a single layer, and then returns the best solution found along the way.
Finally, we implement the critical gap method (CGM)~\cite{serrano2012uncovering,bruno2019community}, which works exactly like the merging heuristic above, but agglomerates the closest nodes first rather than prioritizing moves by how they affect the objective function $Q$.
Details on all these greedy algorithms can be found in Appendix~\ref{appendix:greedy}.

Comparing the modularity $Q^*$ of the partitions found by all these algorithms, we confirm that the DP finds a superior solution in all cases, sometimes by a large margin.
The partitions found by the merge heuristic and the CGM look visually acceptable---and even arguably  better than those found with the DP method!
However, the large-scale structures found by heuristics are mirages produced by an imperfect maximization routine.
The DP algorithm finds solutions of higher quality with small additional layers that account for nodes whose score is uncorrelated with the community structure.
Hence, this first experiment demonstrates that the objective function ought to be re-designed if the solution with more homogeneous and larger layers is the desired outcome---our true maximizer reveals these defects of the objective.

\begin{figure}
  \centering
  \includegraphics[width=0.9\linewidth]{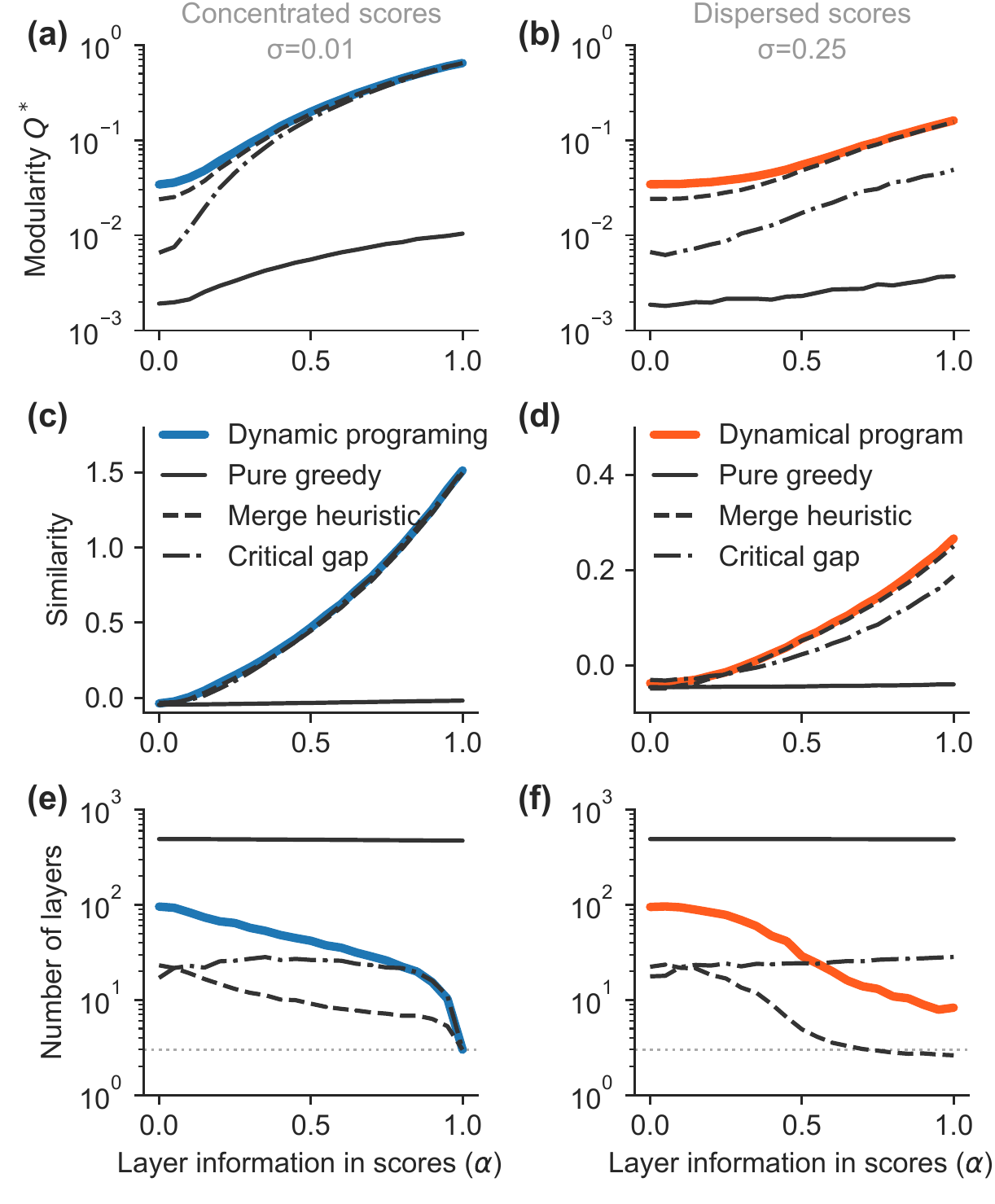}
  \caption{\textbf{Systematic analysis of synthetic networks.}
  We generate networks of $n=500$ nodes with three latent layers and control the probability $\alpha$ that layers' central scores $\mu_\ell\in\{0.25,0.50,0.75\}$ determine a node's score $x_i\in[0,1]$.
  We also control the concentration of the scores around these means, with $\sigma=0.01$ corresponding to very tightly grouped scores (left column) and $\sigma=0.25$ to highly dispersed scores (right column).
  For each network, we run four algorithms aiming to maximize modularity.
  \textbf{(a,b)}. Average modularity of the optimal partition. 
  \textbf{(c,d)}. Average similarity of the optimal partition and ground truth, measured in bits with the reduced mutual information~\cite{newman2020improved}. A reduced mutual information of 0 or less indicates that the partition found by an algorithm contains no information, in the information-theoretic sense, on the planted partition (i.e., no better than randomly guessing).
  \textbf{(e,f)}. Average number of layers in the optimal partition. There are three layers in the ground truth (thin gray line).}
  \label{fig:systematic}
\end{figure}

To systematize these observations, we repeat the experiment above on thousands of networks generated with the same model while varying the amount of layer information contained in the scores (via $\alpha$) and the variance of the scores within layers (via $\sigma$).
Our results are displayed in Fig.~\ref{fig:systematic} and confirm that DP always finds a better partition regardless of the generating parameters, though (some) greedy algorithms come close in some regimes.

Among heuristics, the merging heuristic is the best and notably outperforms the critical gap method.
As mentioned above, both methods work essentially in the same way, but the greedy merging algorithm prioritizes moves by their impact on the objective $Q$, while the CGM only prioritizes good merges indirectly, by relying on the observation that closeby nodes often belong to the same community.
This turns out to be a worse strategy, as can be seen, for example, in Fig.~\ref{fig:systematic}~(a) and  Fig.~\ref{fig:systematic}~(b), where there is a significant gap between the modularity $Q^*$ of the partition found by these two methods.
Hence, while we always  recommend using the DP algorithm, these results suggest that the merging heuristic should be preferred over the CGM if one insists on using approximate maximization routines.

Finally, we highlight that the purely greedy strategy performs poorly.
It finds low-quality solutions with far too many layers.
Recall that unlike the merging heuristics and the CGM, this algorithm does not accept temporary reductions to the objective function, and instead stops as soon as no improvements can be found.
In one-dimensional networks, the flexibility to accept bad tradeoffs turns out to be particularly critical, as a single outlier node within an otherwise high-quality layer can halt the progress of a greedy method.
A purely greedy method will leave this node in its own layer, and split the surrounding layers as a result.
In contrast, merging heuristics can force such merges at the price of a temporary decrease to the objective, eventually finding the larger-scale structure.
These results should not be taken to discredit purely greedy algorithms in general (e.g., Louvain-style algorithms), but they do show that the technique does not translate well to networks embedded in one dimension.

% ~~~~~~~~~~~~~~~~~~~~~~~~~~~~~~~~~~~~~~~~~~~~~~~
\subsection{Splitting social hierarchies}
\label{subsec:addhealth}
% ~~~~~~~~~~~~~~~~~~~~~~~~~~~~~~~~~~~~~~~~~~~~~~~

Next, we use our methodology to analyze real-world social networks measured as part of the US National Longitudinal Study of Adolescent to Adult Health (the ``AddHealth'' study) in 1994 and 1995.
These networks span $84$ schools attended by between $22$ and $2188$ students ($665$ on average).\footnote{We limit our study to the $81$ schools with well-formed data sets and thereby exclude communities 1, 5, and 48.}
They were constructed with the so-called roster method~\cite{henry2012survey} by asking students to select up to $10$ of their friends from a list, capping the number of nominations to five males and five females at most.
This leads to an average of $3128$ nominations per school across the data set.

In Ref.~\cite{ball2013friendship}, Ball and Newman observe that social hierarchies in these schools are captured by unreciprocated ``aspirational'' nominations, i.e., instances of low-status students listing high-status students as their friends, while not getting nominated back.
They use this observation to sort students in a hierarchy, with highly sought-after students at the top and students with no reciprocated nominations at the bottom.
For most networks, the resulting status hierarchy correlates moderately with schooling year, with an average \emph{absolute} Pearson correlation coefficient of $|r|=0.43$ between the position $x_i$ of students and their grade $g_i\in\{7,8,9,10,11,12\}$.
(A student's status typically rises with grades, but the correlation is strong and inverted in two of these schools.)

We apply the DP optimizer to these hierarchies, treating the status $x_i$ of students as their score, and find partitions in layers that maximize each of the objectives outlined in Eqs.~\eqref{eq:modularity}--\eqref{eq:obj_balance} (we set $\lambda=0.5$ in the ``balanced'' objective of Eq.~\eqref{eq:obj_balance}).

\begin{table}
\centering
\caption{Average similarity of the partitions induced by node attributes and the separation in layers with various objective functions. The similarity is calculated with the reduced mutual information (RMI)~\cite{newman2020improved}, where an RMI of 0 or less indicates that no information is shared between the detected layers and attributes.}
\begin{tabular}{lc|cccc}
\hline
\hline
Objective & Eq. &   Grade &     Race &  School &       Sex \\
\hline
Modularity   & \eqref{eq:modularity}      &  \;\;  0.63    \;\; &  \;\;  0.02     \;\; &  \;\;  0.26  \;\; &  \;\;  $\leq0$ \;\;\\
Egalitarian  & \eqref{eq:obj_egalitarian} &  \;\;  0.54    \;\; &  \;\; $\leq0$   \;\; &  \;\;  0.24  \;\; &  \;\;  $\leq0$ \;\;\\
Dominance    & \eqref{eq:obj_dominance}   &  \;\;  0.07    \;\; &  \;\; $\leq0$   \;\; &  \;\;  0.10  \;\; &  \;\;  $\leq0$ \;\;\\
Balanced     & \eqref{eq:obj_balance}     &  \;\;  0.26    \;\; &  \;\; $\leq0$   \;\; &  \;\;  0.16  \;\; &  \;\; $\leq0$ \;\;\\
\hline 
\hline
\end{tabular}
\label{table:add_health}
\end{table}

Table~\ref{table:add_health} shows the similarity of the partitions found by the DP and the ones induced by the demographics of students, namely their grades (between 7 and 12), race (Asian, Black, Hispanic, mixed/others, unreported, or White), sex (male, female, unreported), and school (some datasets combine two schools with cross-school nominations). 
These similarity scores are averaged over all 81 schools.

The layers found with 3 out of 4 objectives are, on average, most similar to the partitions induced by splitting students by school years (grades).
We find no evidence that social strata contain  information on the partition in race or sex, indicating that these demographics are not concentrated in any particular strata.
Finally, we see that the layers contain some information about schools, which is expected since communities with more than one school consist of a high school and a ``feeder'' middle school~\cite{ball2013friendship}---such that the school variable is itself correlated with grades.

Not shown in the table is the fact that we always find a finer separation in layers than what can be explained by the students' demographics alone.
The optimal partitions comprise $58.2$ layers on average when using the ``egalitarian'' objective, of $50.0$ layers for modularity, $131.0$ for the ``balanced'' objective, and $164.3$ (!) for the ``dominance'' objective. 
These findings can be better understood in light of our previous results with the planted partition model (c.f. Sec.~\ref{subsec:results_simulation_study}).
Networks are complex and high-dimensional objects, and a single explanatory variable may easily misplace nodes.
As we have seen in simulations, the optimal partition will often include additional layers to account for these nodes.
In the context of network data analysis, our recommendation is thus to focus on designing objectives robust to misplaced nodes or exploring alternative embedding methods that limit such ``mistakes.''

% ~~~~~~~~~~~~~~~~~~~~~~~~~~~~~~~~~~~~~~~~~~~~~~~
\subsection{Communities in hyperbolic embeddings}
% ~~~~~~~~~~~~~~~~~~~~~~~~~~~~~~~~~~~~~~~~~~~~~~~

\begin{figure}
    \centering
  \includegraphics[width=\linewidth, trim=1cm 0cm 1cm 0cm, clip=true]{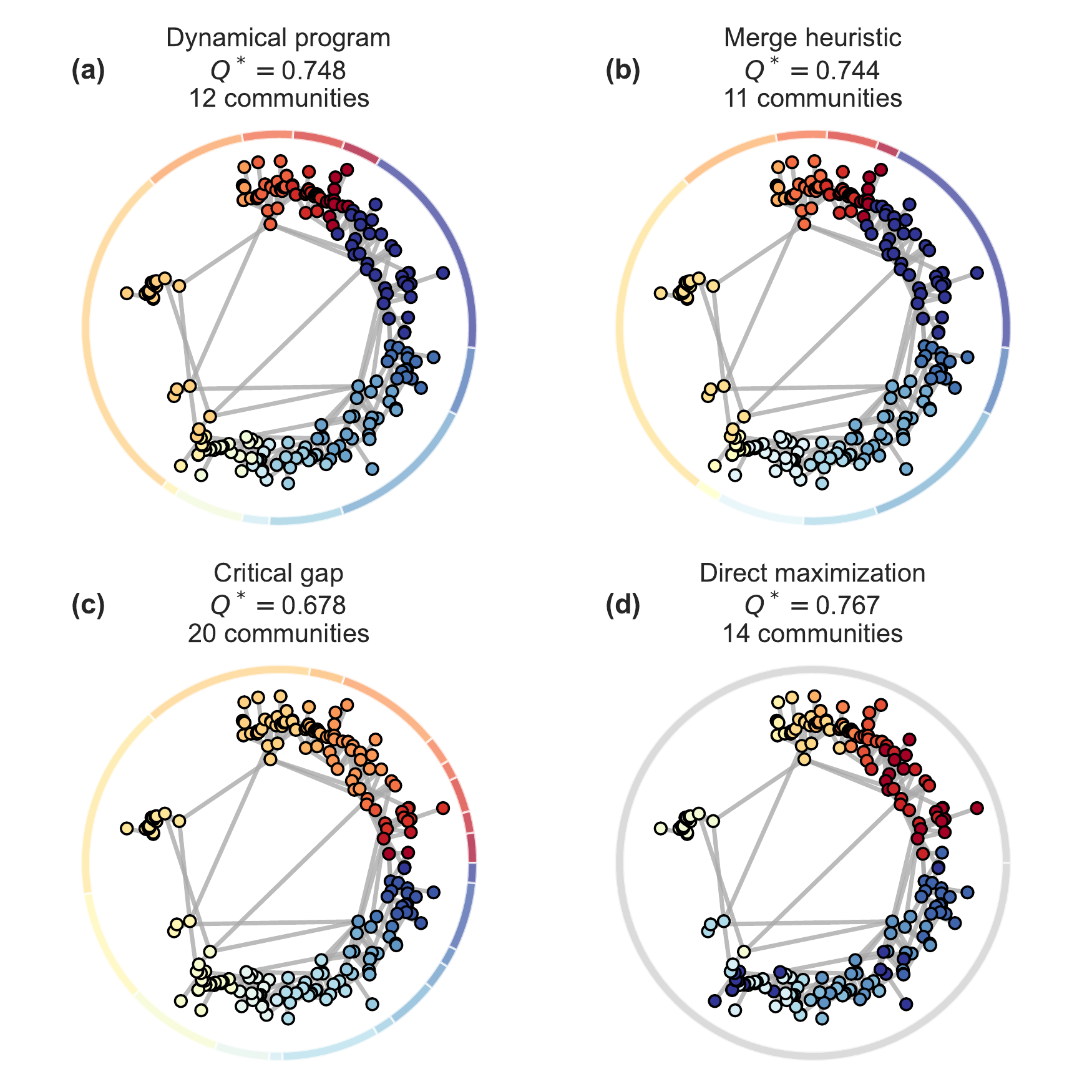}
  \caption{\textbf{Community detection with hyperbolic embeddings}. Analysis of the Tokyo subways network~\cite{roth2012long} embedded in $\mathbb{H}^2$ with the \texttt{Mercator} algorithm.  (a) Communities retrieved by clustering nodes along the angular dimension with dynamic programming. Both the outer circle and the node colors denote communities. (b) Communities found by the merge heuristic. (c) Communities found by the critical gap method. (d) Communities found by directly maximizing the objective~\cite{newman2004finding}, without using the embedding as an intermediate step.
  In all cases, we use a modularity objective with the configuration model as the null model; see Eqs.~\eqref{eq:modularity} and~\eqref{eq:cm_null}.}
  \label{fig:viz_hyperbolic}
\end{figure}

For our final case study, we identify the large-scale structure of over 200 networks with a two steps process, namely an embedding step to estimate the positions of nodes in a latent space, followed by a clustering of close-by nodes~\cite{chami2019hyperbolic,boguna2021network,hoff2002latent}.
Unlike in the previous case study, we use a domain-agnostic embedding with no direct theoretical interpretation; instead, we embed the nodes in the two-dimensional hyperbolic space $\mathbb{H}^2$.
The positions of nodes in this space are typically interpreted as the ``popularity'' $r_i$ along the radial dimension (nodes closer to the center of the disk are more popular) and as the ``similarity'' along the angular dimension (nodes with similar angles $x_i$ are alike)~\cite{boguna2021network}.
Recent work has shown that clustering the nodes along the angular dimensions can uncover large-scale structures in networks~\cite{bruno2019community,tandon2021community}.
Thus far, this clustering has been conducted with heuristics such as the critical gap method (CGM)~\cite{serrano2012uncovering}.
Our proposed algorithm can be used in its place since the angular dimension is a one-dimensional space with periodic boundary conditions.
The groups we have called ``layers'' until now can be thought of ``\emph{communities}'' in this context~\cite{fortunato2010community}.

To form an intuition for the behavior of this embedding-based detection algorithm, we first apply it to a network of subway stations~\cite{roth2012long} embedded in $\mathbb{H}^2$ with the \texttt{Mercator} algorithm~\cite{garcia2019mercator}.
Nodes in this network denote subway stations, and edges denote lines running between stations.
Angular positions $x_i\in[0,2\pi]$ are found by maximizing the likelihood of the network under a hyperbolic network model.
We expect the embedding to reveal meaningful structure because this subway network is physically constrained in the first place~\cite{tandon2021community}.
Once we have obtained the position of the nodes, we use the DP to maximize a modularity objective, as defined in Eqs.~\eqref{eq:modularity} and~\eqref{eq:cm_null}.

The results of this detection procedure are shown in Fig.~\ref{fig:viz_hyperbolic} and are compared against the partition found with the more typical CGM.
We also show results found with different heuristic methods adapted from the previous sections and the communities found without using any embedding at all, which we refer to as the ``direct maximization'' result~\cite{newman2004finding}.
As can be seen from the result, the DP performs better than other embedding-based methods and finds fewer communities than the CGM in this particular case.
The merging heuristic is again superior to the CGM since it improves on the objective more directly than the latter.
And notably, the best partition derived from the embedding does not surpass the partition found by direct maximization; see Appendix~\ref{appendix:optimality} for additional discussion of this point.

\begin{figure}
  \centering
  \includegraphics[width=\linewidth]{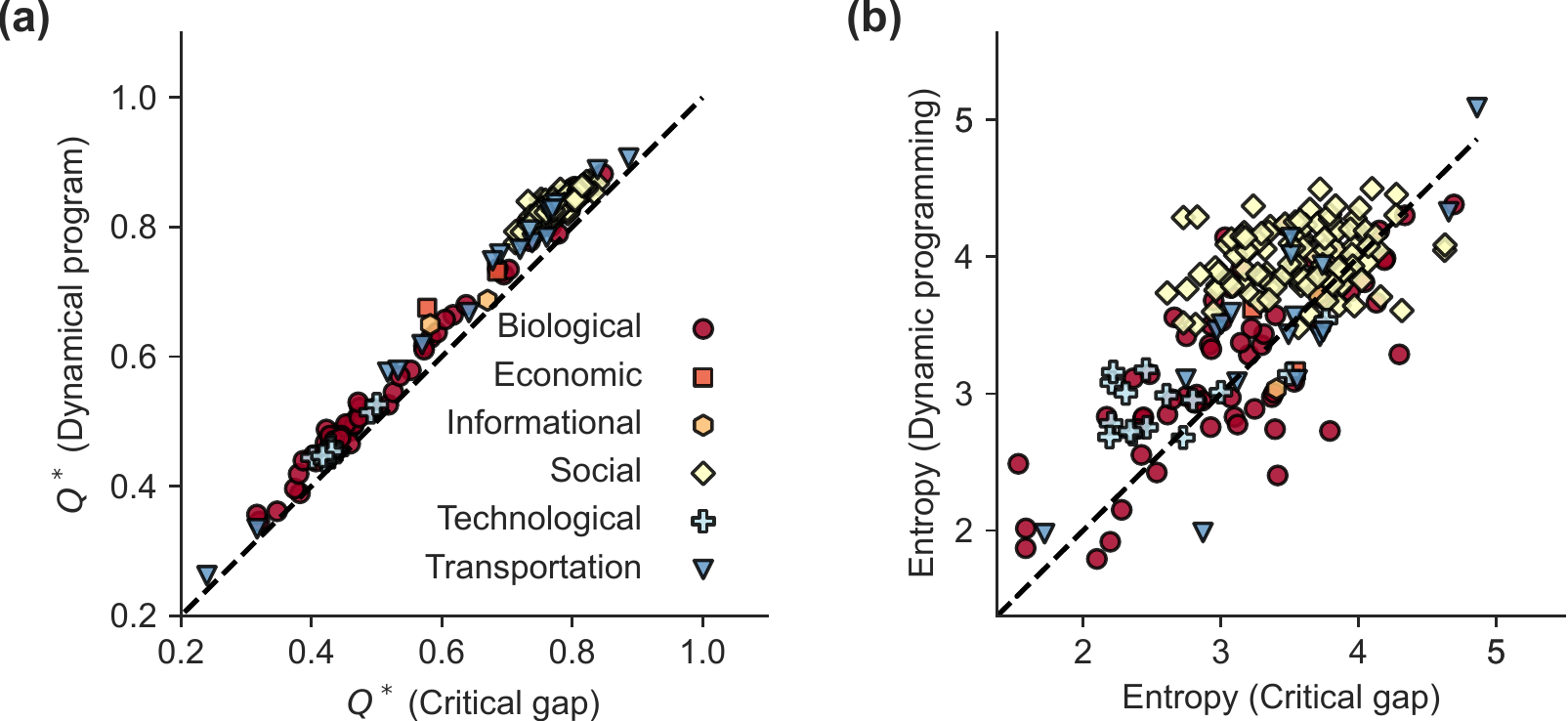}
  \caption{\textbf{Exact and approximate detection methods for hyperbolic embeddings}. We compare the  dynamic programming and critical gap methods on a subset of 211 networks of the \texttt{CommunityFitNet} benchmark~\cite{ghasemian2018evaluating}, embedded in hyperbolic space with the \texttt{Mercator} algorithm~\cite{garcia2019mercator}.
  (a) Normalized modularity of the partitions, obtained by dividing Eq.~\eqref{eq:modularity} by the number of edges in each network.
  (b) Normalized entropy of the distribution of community sizes (in bits).
  The diagonal line is shown in both panels; markers are on this diagonal whenever the two algorithms return identical partitions.}
  \label{fig:exact_vs_cgm_hyperbolic}
\end{figure}

We also conduct a more systematic analysis of our methods using a subset of networks taken from the \texttt{CommunityFitNet} benchmark~\cite{ghasemian2018evaluating}.
We exclude bipartite and directed networks, for which the \texttt{Mercator} algorithm is undefined, as well as networks with fewer than 200 nodes.
The outcomes of this experiment are summarized in Fig.~\ref{fig:exact_vs_cgm_hyperbolic}, where we show results obtained by maximizing the modularity with the standard configuration model as the null model.
As expected, the DP-based algorithm always finds a better optimum, though the CGM often comes quite close (Pearson correlation coefficient of $r=0.995$).
However, we find important differences upon close inspection of the returned partitions.
For instance, the CGM finds $1.16\pm0.60$ times more communities than the optimal solution (mean, standard deviation), and the correlation between the number of communities found by the two methods is only $r=0.685$.
The composition of these communities also differs significantly, and this can be quantified with the entropy of the distribution of community sizes, defined as
\begin{equation}
  S = -\sum_{\ell=1}^q \frac{n_\ell}{n} \log_2 \frac{n_\ell}{n}
\end{equation}
where $n_\ell$ is the number of nodes in community $\ell$.
This entropy is large when communities have equal sizes and small when nodes concentrate in only a few communities.
Figure~\ref{fig:exact_vs_cgm_hyperbolic}~(b) shows how the entropy of the optimal partition is affected by the choice of algorithm and demonstrates that the DP finds communities that are more balanced on average, with an average ratio of entropy of $S_{\mathrm{DP}}/S_{\mathrm{CGM}} = 1.096 \pm 0.15$ for the DP and CGM.
These two quantities are only moderately correlated within our sample ($r=0.707$).

In sum, this systematic comparison sheds light on the potential pitfall of using the CGM in that it can uncover partitions with almost optimal modularity but whose composition may differ greatly from the optimal partition (obtained using DP).

% ~~~~~~~~~~~~~~~~~~~~~~~~~~~~~~~~~~~~~~~~~~~~~~~
% ~~~~~~~~~~~~~~~~~~~~~~~~~~~~~~~~~~~~~~~~~~~~~~~
\section{Conclusion}
\label{sec:conclusion}
% ~~~~~~~~~~~~~~~~~~~~~~~~~~~~~~~~~~~~~~~~~~~~~~~
% ~~~~~~~~~~~~~~~~~~~~~~~~~~~~~~~~~~~~~~~~~~~~~~~

This paper presented a clustering algorithm for networks embedded in one-dimensional spaces, such as the real line or the circle.
The algorithm uses well-known dynamic programming (DP) ideas~\cite{jackson2005algorithm} to achieve optimal clustering results in polynomial time.
As such, it improves upon the heuristics used to solve this problem in network science, including greedy strategies, merge heuristics, and the critical gap method, which all return suboptimal results in a similar amount of time.

Nonetheless, the DP algorithm has limitations that could be improved in future research.
First, the quadratic scaling of its complexity with $n$ is too rapid for use with very large networks.
Various strategies could alleviate this scaling issue.
For example, it might be possible to adopt a divide-and-conquer strategy and split the network into chunks of less-than-linear size that could then be processed independently.
This would lead to better scaling, at the cost of some accuracy.
One could also perhaps find optimization opportunities within the maximization step in Eq.~\eqref{eq:best_split_quality}---say, with a clever search strategy or by using special-purpose hardware, like the \texttt{DPX} instruction set for dynamic programming on graphics processing units~\cite{choquette2023nvidia}.
The payoff could be large, as this step is in the inner loop of the algorithm.
Second, our generalization of the DP strategy to embedding on the circle uses a potentially costly search step; see Appendix.~\ref{appendix:periodic}.
Can periodic boundary conditions be handled more elegantly? And most importantly, more cheaply?
Third, as discussed in Sec.~\ref{sec:problem}, not all objective functions can be optimized with this DP strategy.
An improved algorithm able to fit generic objective functions without significant overhead would be a valuable addition to the network science algorithmic toolkit.
And fourth, the DP strategy is limited to linear clustering, because the recursive structure of Eq.~\eqref{eq:best_split_quality} breaks down in more than one dimension.
Can something clever be done to accommodate embedding in higher dimensions? 
Machine learning methods operate in hundreds if not thousands of dimensions and providing provably optimal clustering could improve the whole machine learning pipeline.
Perhaps the DP primitive could be applied to each dimension, before aggregating the results.
Or alternatively, we may envision direct a generalization of Eq.~\eqref{eq:best_split_quality} to high-dimensional spaces obtained by clearly defining contiguity in $\mathbb{R}^d$.

Despite these limitations, our DP solution already has the potential to facilitate research on several problems of interest to network science. 
First, with optimal clustering out of the way, we can now turn our attention to the fundamental question of choosing an objective function whose principles align with a paired embedding method.
For example, if one wishes to retrieve social strata using embedding and the bespoke objectives discussed in Secs.~\ref{sec:objective} and \ref{subsec:addhealth}, should the scores be estimated with the complex statistical model of Ball and Newman~\cite{ball2013friendship}, or is Pagerank sufficient, or even better?
And in general, which pairings work best?
Large-scale simulation studies can now be used to look into this issue.
Second, we are now free to build complex pipelines on top of the DP primitive, for instance, to quantify the effect of sampling errors on clustering.
This could be done by subsampling a network before or after embedding and studying the impact on clusters.
Third and most importantly, scientific questions that involve hierarchies and clusters are now within reach~\cite{de2018physical}.
Our DP algorithm thus paves the way to a more thorough understanding of the relationship between these embedding and large-scale emergent structures in networks.

\section*{Acknowledgments}
This work was supported in part by the James S McDonnell Foundation (JGY), the Fonds de recherche du Qu\'ebec -- Nature et technologies (AA), the Natural Sciences and Engineering Research Council of Canada (AA), and the Sentinelle Nord program of Universit\'e Laval, funded by the Canada First Research Excellence Fund (AA).
We thank Eleanor Power, Elizabeth Bruch, Caterina De Bacco, Johan Ugander, and George T. Cantwell for helpful discussions.\\

This research uses data from Add Health, a program project directed by Kathleen Mullan Harris and designed by J. Richard Udry, Peter S. Bearman, and Kathleen Mullan Harris at the University of North Carolina at Chapel Hill, and funded by grant P01-HD31921 from the Eunice Kennedy Shriver National Institute of Child Health and Human Development, with cooperative funding from 23 other federal agencies and foundations. Information on obtaining the Add Health data files is available on the Add Health website (\url{https://addhealth.cpc.unc.edu}). No direct support was received from grant P01-HD31921 for this analysis.
\appendix
\section*{Appendix}

% =========================
\section{Complexity and implementation}
\label{appendix:improved_implementation}
% =========================
In this Appendix, we analyze the complexity of the dynamic programming algorithm discussed in Sec.~\ref{sec:method}, and describe a few optimization opportunities.

The algorithm's most costly step is in the inner loop, namely the maximization in Eq.~\eqref{eq:best_split_quality}.
Each evaluation of Eq.~\eqref{eq:best_split_quality} requires $j$ calls to the layer-quality function $f(\cdot)$, meaning that a complete execution of the algorithm visits this steps $j$ times for $j=1,...,n$.
The layer quality is thus called $\sum_{j=1}^n j = \binom{n+1}{2}$ times, which implies an overall time-complexity that scales at least as $O(n^2)$ with the number of nodes $n$.
This bound is achieved if f$(\cdot)$ can be evaluated in constant time, but when calls to $f(\cdot)$ depend on $n$, the total running time may increase faster than $O(n^2)$.
For example, if it takes at least as many operations as there are nodes in a layer to evaluate $f(\cdot)$, then the number of operations will be lower bounded by
\begin{equation}
  \label{eq:bad_scaling}
  \sum_{j=1}^{n} \sum_{k=1}^j (j - k)  = \binom{n+1}{3} = O(n^3),
\end{equation}
with even steeper scaling for, say, quality functions whose complexity grows quadratically with the number of nodes in each layer.

It is possible to tame this rapid growth by choosing layer-quality functions carefully.
One possibility is to restrict our choice to any function that can be written as
\begin{equation}
  \label{eq:separable_condition}
  f(L_{k,j}) = \sum_{u=k}^{j}\sum_{v=u}^j f_{uv},
\end{equation}
i.e., as a sum over the ``quality increments'' $f_{uv}$  of putting a pair of nodes $(u,v)$ in the same layer.
Modularity-like objectives can be written in this way, and so can partitioning objectives and many other so-called ``pairwise'' models~\cite{Young2018}.

With this type of quality function, we can calculate the quality of a layer as
\begin{multline}
  \label{eq:update_equation}
  f(L_{k,j}) \\= f(L_{k, j- 1}) + f(L_{k+1, j})  - f(L_{k+1,j-1}) + f_{kj}
\end{multline}
by  expressing $f(L_{k,j})$ as the sum of two overlapping sets minus their intersection, with an extra added term $f_{kj}$ to account for the interaction of nodes at the two extremes of the layer.
This observation allows us to evaluate a layer's quality in $O(1)$ steps since we can execute all calculations in an order guaranteeing that all quantities on the right-hand-side of Eq.~\eqref{eq:update_equation} are already known by the time we have to evaluate the left-hand-side.
More precisely, we 
\begin{enumerate}
   \item  execute the maximization in Eq.~\eqref{eq:best_split_quality} in decreasing order of $k=j-1,j-2,...,1$; this allows us to compute $f(L_{k+1, j}) $ before we need to compute $ f(L_{k,j})$.
  \item builds the recursion from the bottom up, starting at $j=1$ and moving upward; this ensures that  $f(L_{k, j- 1})$ and $f(L_{k+1,j-1})$ are evaluated before $(L_{k,j})$ because these terms only involve the first $j-1$ nodes.
\end{enumerate}
We store these quantities in an array of layer qualities indexed by their limits $(k,j)$ when they are first computed, which allows for later retrieval in $O(1)$ time.

This modification yields an overall algorithm that runs in $O(n^2)$ steps since we can replace $j-k$ by a constant in the complexity calculation of Eq.~\eqref{eq:bad_scaling}.

Different types of objective functions can lead to similar update equations in the spirit of Eq.~\eqref{eq:update_equation}; for example, a symmetric sum (notice the summation indices):
\begin{equation*}
   f(L_{k,j}) = \sum_{u=k}^j \sum_{v=k}^{j} f_{uv},
\end{equation*}
would yield essentially the same equation as Eq.~\eqref{eq:update_equation}, with an additional term $f_{jk}$.

% =========================
\section{Modification for periodic conditions}
\label{appendix:periodic}
% =========================

In this Appendix, we describe the modifications needed for the algorithm to work when the embedding space is periodic.

The periodic problem is slightly harder since a periodic partition with $q$ layers can be described using $q$ cut points instead $q-1$, the number of cut points needed in the non-periodic case.
The total number of partitions in exactly $q$ layers is thus
\begin{equation}
  \Omega(n, q) = \binom{n}{q} = O(n^q)
\end{equation}
and the total number of possible partitions is now $\sum_{q=0}^n = 2^n$ instead of $ 2^{n-1}$.

We handle this larger search space by treating the DP algorithm as a computational primitive.
Our strategy is to unfold the periodic embedding next to some node $u$, such that nodes $u$ and $u+1$ are no longer next to each other in the embedding space.
This forces a cut at $u$ and creates a linear embedding of the nodes, on which we can run the DP to obtain the remaining layers.
We denote the resulting set of layers as $\mathcal{L}^*_u$ to emphasize its dependency upon $u$.
(The DP can, in principle, return a single layer, in which case the cut at $u$ has no effect---all nodes are in the same layer.)
 
\begin{figure}
  \centering
  \includegraphics[width=\linewidth]{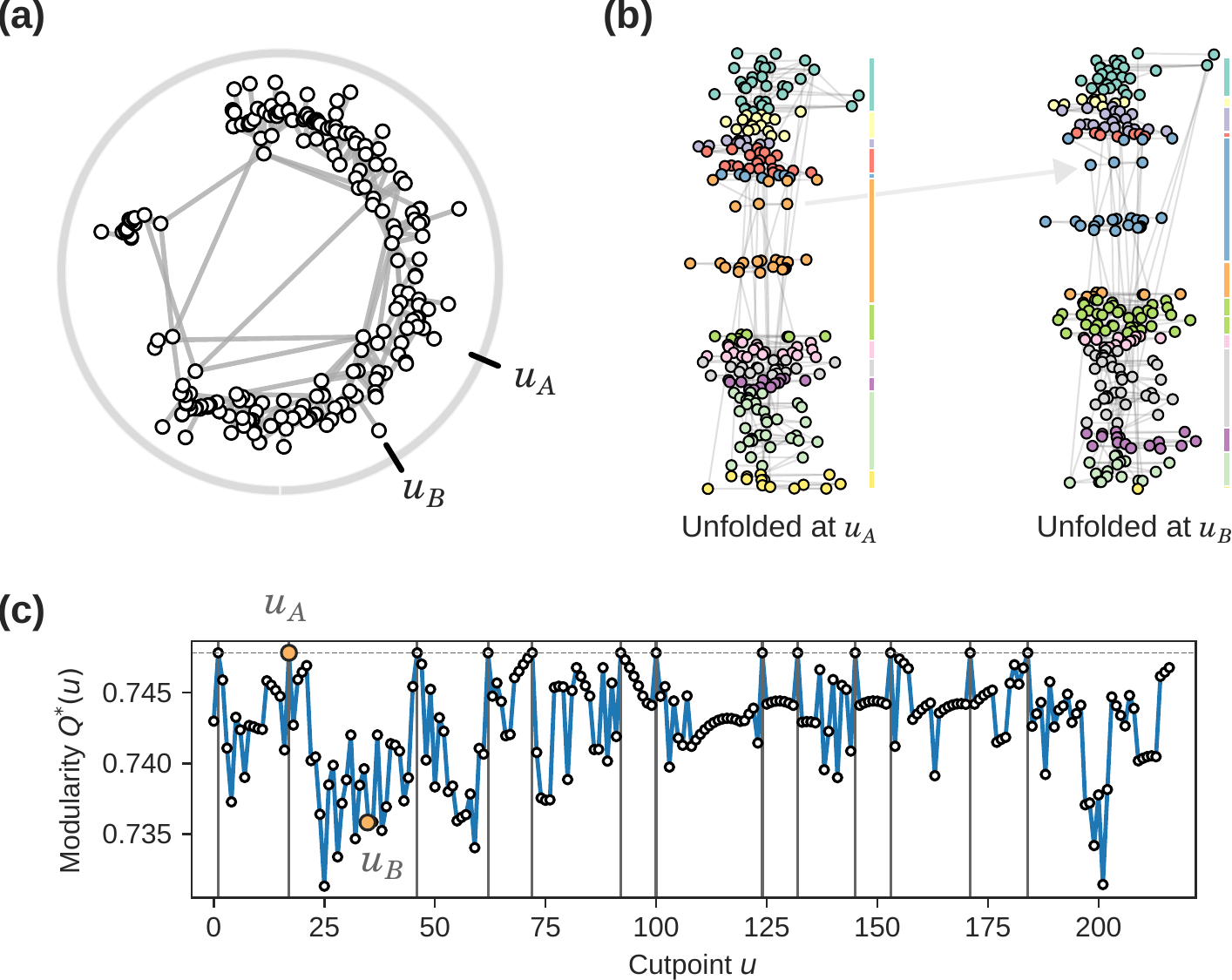}
  \caption{\textbf{Layers of a network embedded in a periodic space}. \textbf{(a)} Periodic embedding of the subway network shown in Fig.~\ref{fig:viz_hyperbolic}. We highlight two potential cutpoints $u_A$ and $u_B$.
  \textbf{(b)} Two linear embeddings obtained by unfolding the periodic embedding between $u_A$ and $u_A+1$, and $u_B$ and $u_B+1$ respectively. The graphs are the same, but the positions of the nodes differ. 
  As a result, a few long-range ties appear in the embedding obtained by unfolding at $u_B$, and the optimal layers also change.
  \textbf{(c)} We find a globally optimal set of layers by searching through cutpoints $u$.
  There are several optima, here marked with vertical lines.}
  \label{fig:unfolding}
\end{figure}

Different choices of $u$ will generally lead to optima with different qualities $Q^*(u):=Q(\mathcal{L}^*_u)$, and the quality  need not be a well-behaved function of $u$, see Fig.~\ref{fig:unfolding}~(c) for an example.
Hence, to find the true global optimum, we solve a search over choices of cutpoints $u$.

The simplest method for finding a true global optimum $\mathcal{L}^*$ is an exhaustive search through the complete set of partitions $\mathcal{C}=\{\mathcal{L}^*_u\}_{u=1,..,n}$.
This search has a linear cost in the number of nodes $n$, which leads to an overall complexity of $O\bigl(n\times f(n)\bigr)$ where $f(n)$ is the complexity of the DP algorithm.
The resulting algorithm is manageable but relatively slow.

We can do better by noticing that there is more than one solution to the search problem because the embedding is periodic.
For example, Fig.~\ref{fig:unfolding}~(c) shows a network for which 13 different cutpoints lead to solutions of identical modularity.
This is due to the fact that imposing a cut at any of the natural cutpoints $u$ of the global optimum will yield an optimal solution.

To make use of this observation, we introduce the notion of a stable partition $\mathcal{S}$.
A partition is stable if we cannot find a solution of higher quality $Q$ by forcing a cut between any of the layers of $\mathcal{S}$ and running the DP on the resulting linear embedding.
Three properties of stable partition are important to highlight.
First, two stable partitions of different quality cannot share any cut points---otherwise, one would be able to find an improvement over either by splitting the embedding at that cutpoint.
Second, an \emph{un}stable partition may share cut points with a stable partition.
And third, the global optimum is stable by definition.

The properties of stable partitions suggest a simple greedy search algorithm over cutpoints $u$ since unstable partitions are connected to stable partitions by the cutting operation, and the global optimal is stable under this operation.
Starting from an arbitrary node $u$, we unfold the embedding and obtain $\mathcal{L}_u^*$.
We then iterate through the layers of $\mathcal{L}^*_u$, unfold the embedding at every node $u'$ on the boundary of these layers, and run the DP on each.
For a constant number of layers, this test is of no greater complexity than running the DP.
If $\mathcal{L}^*_{u'}$ is of superior quality, then $\mathcal{L}^*_u$ is not stable, but the new partition might be.
So we add  $u'$ to a list of potential optimal cutpoints for all such $u'$ and repeat the procedure (while keeping track of already attempted solutions to avoid infinite loops).
In practice, we find that we can reach the optimum partitions by repeating these steps a handful of times, most often once or twice, because most unstable partitions still contain one of the cutpoints of the global optimum.
As long as the number of iterations needed to reach a stable partition does not scale too rapidly with $n$, this strategy thus provides a fast way to the optimum.

% ==============================
\section{Greedy algorithms}
\label{appendix:greedy}
% ==============================
\begin{figure}
  \centering
  \includegraphics[width=0.6\linewidth]{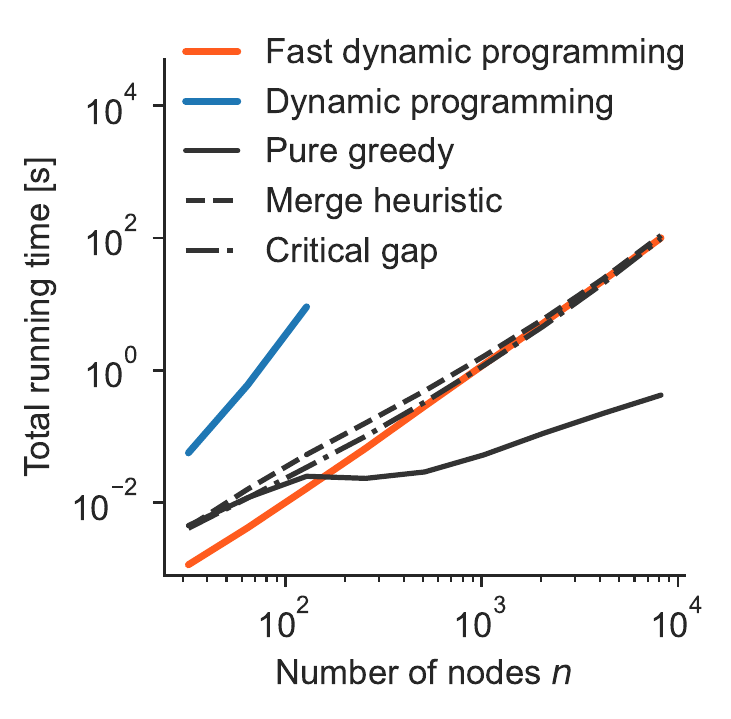}
  \caption{\textbf{Running time of all algorithms as a function of network size.}
  We apply all algorithms to the test networks described in Sec.~\ref{subsec:results_simulation_study} while varying the number of nodes $n$.
  We fix the number of (equal-sized) layers to $q=3$ communities and adjust the connection probabilities to keep the network sparse, with an average degree of $10$ and a fixed ratio $\epsilon=\langle k_{\mathrm{out}}\rangle / \langle k_{\mathrm{in}}\rangle=0.2$ of the external and internal average degree of nodes, given by $\langle k_{\mathrm{out}}\rangle = p_{\mathrm{out}} (q-1)n/q$ and $\langle k_{\mathrm{in}}\rangle = p_{\mathrm{in}} n/q$, respectively.
  The other parameters are $\alpha=0.9$, $\mu_\ell\in\{0.25,0.50,0.75\}$, and $\sigma = 0.25$.
  The fast dynamic programming approach uses the optimizations discussed in Appendix~\ref{appendix:improved_implementation} (solid orange line), while the simple dynamic program implements Eq.~\eqref{eq:best_split_quality} directly (solid blue line).}
  \label{fig:scaling}
\end{figure}

This Appendix provides details on the greedy algorithms used in the main text.
Figure~\ref{fig:scaling} compares their total running time on synthetic networks.
The DP algorithm and all greedy algorithms run in minutes on networks of 8192 nodes, using a \texttt{python} implementation of the algorithms on a standard latop.
In comparison, the brute-force search runs in about an hour on a network of 32 nodes.
Solution quality is analyzed in the main text, see Fig.~\ref{fig:systematic}.
\\

\noindent\textbf{Purely greedy strategy}. 
This method is a loose adaptation of the Louvain algorithm~\cite{blondel2008fast} to a network embedded in one dimension.

We initialize each node in its own layer.
We then visit every node in a random order and check whether moving it to the layer of the node directly above or below would increase the objective function. 
Of these two proposed changes, we pick the one that most increases the objective. 
We do nothing if neither lead to an improvement.
We repeat this process until a local optimum is reached. 

This algorithm can be implemented efficiently since all the moves only require local information.
\\

\noindent\textbf{Merge heuristics}. 
This method adapts the fast unfolding algorithm of Ref.~\cite{clauset2004finding} to one-dimensional problems.

The algorithm proceeds by merging adjacent layers, thereby reducing the number of layers at each step.
We initialize each node in its own layer, and compute the update $\Delta Q_{rs}$ one would obtain by merging adjacent layers $r$ and $s$, for all $(r,s)$. 
We then apply the merge $(r^*, s^*) \to r'$ that most increases $Q$--or the least negative one if none can improve $Q$---and update $\Delta Q_{r'-1, r'}, \Delta_{r', r'+1}$ to reflect this merge.
We repeat this process until all nodes are in a single layer and return the best partition found along the way.

This algorithm can be implemented efficiently by storing merges on a heap for fast retrieval.
Further, very few heap updates are necessary during execution since $\Delta Q_{rs}$ changes for only $2$ pairs of layers at every strep.\\

\noindent\textbf{Critical gap method}. 
This method is directly adapted from Ref.~\cite{bruno2019community}.
It resembles the merge heuristic above but prioritizes merges based on node distances rather than changes to the objective function.
More precisely, the algorithm computes a gap $d_{rs}$, defined as the distance between nodes on either end of two adjacent layers $r$ and $s$, and performs merges in order of increasing gap.
Again, we apply all merges even if they decrease the objective $Q$, and return the best of all encountered partitions.

This algorithm can be implemented using the architecture of the merge heuristic by precomputing all gap distances and storing them on a heap.
While the critical gap method may look more efficient than the merge heuristics because it operates on gap widths $d_{rs}$ (cheap to compute) instead of changes to the objective $\Delta Q_{rs}$ (expensive to maintain), it is in fact just as costly to execute.
This is because one must evaluate the quality $Q$ of each intermediary solution to find the global optimum.
For the class of additive objectives considered in this paper, an efficient way to compute $Q$ is to maintain and update an array of increment $\Delta Q_{rs}$ for each intermediary solution and update $Q$ as layers are merged---just like in the merge heuristics.
Hence, the asymptotic complexity of the two methods is identical.

\section{On the sub-optimality of ultra low-dimensional embedding for community detection}
\label{appendix:optimality}
\begin{figure*}
  \centering
  \includegraphics[width=0.8\linewidth]{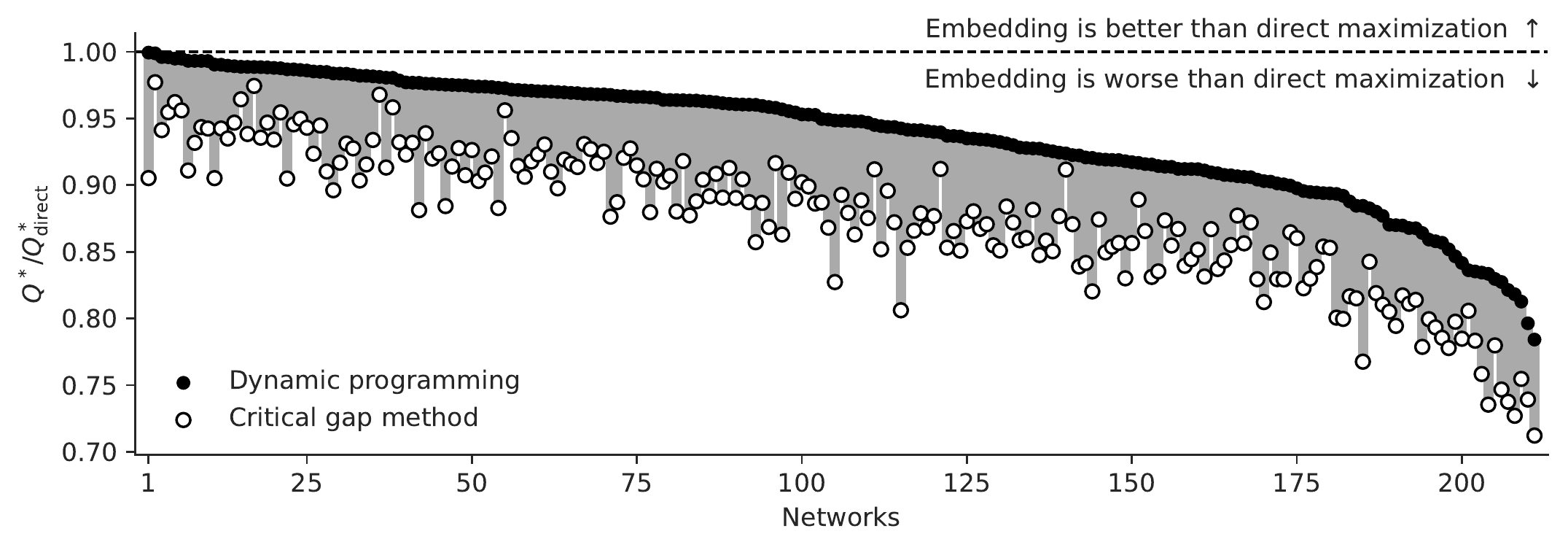}
  \caption{\textbf{Direct maximization versus indirect maximization of the modularity}. Comparison of the modularity $Q_{\mathrm{direct}}^*$ of partitions found by direct maximization~\cite{newman2004finding,ghasemian2018evaluating} and the modularity $Q^*$ of partitions found with dynamic programming or with the critical gap method, given an embedding of the nodes in $\mathbb{H}^2$~\cite{garcia2019mercator}.
  A dotted line denotes the parity for direct and indirect maximization techniques.
  We use the subset of the \texttt{CommunityFitNet}~\cite{ghasemian2018evaluating} benchmark described in the main text, ordered by the quality $Q^*$ of DP partitions.
  }
  \label{supp_fig:direct_vs_exact}
\end{figure*}

The quality of partitions found with embedding methods depends on three key components: (1) the embedding itself, (2) a search strategy over  partitions guided by that embedding, (3) the objective function used to rank solutions.
As shown in the main text, the dynamic programming (DP) algorithm solves step 2 optimally and can thus always improve a community detection pipeline that relies on a one-dimensional embedding.

However, for this to be a viable community detection strategy, the embedding in step 1 needs to capture community structure clearly.
We investigate whether that is the case in  Fig.~\ref{supp_fig:direct_vs_exact}, where we embed networks in $\mathbb{H}^2$ and then cluster nodes along the angular dimension, using a modularity objective and two algorithms: the optimal DP solution and the standard critical gap method.
Crucially, we normalize these results by the modularity $Q_{\mathrm{direct}}$ of the partitions found with a direct (but imperfect) algorithm that does not use an intervening embedding step~\cite{newman2004finding}.

The experiment shows that even a perfect clustering of the embedded nodes is insufficient to outperform direct maximization.
When we use the critical gap method in step 2, we cannot match the modularity of the direct maximization solution for a single network.
Switching to the DP algorithm improves the results markedly, but we only achieve modularity scores within 1\% of those obtained by direct maximization in 12 out of 211 networks, and modularity scores of equal quality in only 2 cases.

We can understand these results in light of recent work showing that the structure of networks is difficult to capture with ultra-low-dimensional representations~\cite{desy2023dimension,jankowski2023dmercator,seshadhri2020impossibility}. 
For example, the member of a social group might be similar in some ways (e.g., shared interest) and dissimilar in other ways (e.g., different socioeconomic status).
An embedding in one dimension is not well-suited to simultaneously reflect both of these facts.
As a result, a clustering of  then odes based on this embedding invariably misses some important large-scale structures.
Direct optimization heuristics, on the other hand, are not constrained and can identify arbitrary communities. 
Our results suggest that the flexibility afforded by working without a one-dimensional  embedding is enough to offset these heuristics' lack of optimality guarantees.

\end{document}